\documentclass[12pt]{article}
\usepackage{multicol}
\usepackage{amssymb}
\usepackage{amsfonts,amssymb,amsmath,amsthm}
\usepackage{epsfig}
\usepackage{slashed}
\usepackage{graphicx}
\usepackage{subfig}
\usepackage{mathrsfs}
\usepackage{bbm}
\usepackage{cite}
\usepackage{enumerate}
\usepackage{slashbox}
\usepackage{color}
\usepackage[colorlinks,linkcolor=red,anchorcolor=red,citecolor=green]{hyperref}
\begin{document}
\renewcommand{\thefootnote}{\fnsymbol{footnote}}
\begin{titlepage}

\vspace{10mm}
\begin{center}
{\Large\bf Thermodynamics of noncommutative high-dimensional AdS black holes with non-Gaussian smeared matter distributions}
\vspace{16mm}

{{\large Yan-Gang Miao${}^{1,2,3,}$\footnote{\em E-mail: miaoyg@nankai.edu.cn}
and Zhen-Ming Xu${}^{1,}$}\footnote{\em E-mail: xuzhenm@mail.nankai.edu.cn}\\

\vspace{6mm}
${}^{1}${\em School of Physics, Nankai University, Tianjin 300071, China}

\vspace{3mm}
${}^{2}${\em State Key Laboratory of Theoretical Physics, Institute of Theoretical Physics, \\Chinese Academy of Sciences, P.O. Box 2735, Beijing 100190, China}

\vspace{3mm}
${}^{3}${\normalsize \em CERN, PH-TH Division,
1211 Geneva 23, Switzerland}
}

\end{center}

\vspace{10mm}
\centerline{{\bf{Abstract}}}
\vspace{6mm}
Considering non-Gaussian smeared matter distributions, we investigate thermodynamic behaviors of the noncommutative high-dimensional Schwarzschild-Tangherlini anti-de Sitter black hole, and obtain the condition for the existence of extreme black holes. We indicate that the Gaussian smeared matter distribution, which is a special case of non-Gaussian smeared matter distributions, is not applicable for the 6- and higher-dimensional black holes due to the hoop conjecture. In particular, the phase transition is analyzed in detail. Moreover, we point out that the Maxwell equal area law maintains for the noncommutative black hole whose Hawking temperature is within a specific range, but fails for that whose the Hawking temperature is beyond this range.

\vskip 20pt
\noindent
{\bf PACS Number(s)}: 04.50.Gh; 04.60.Bc; 04.70.Dy 


\end{titlepage}

\newpage
\renewcommand{\thefootnote}{\arabic{footnote}}
\setcounter{footnote}{0}
\setcounter{page}{2}
\pagenumbering{arabic}

\section{Introduction}
The first published paper on noncommutative spacetime was done by Snyder~\cite{HS} for the purpose to cure ultraviolate divergences in quantum field theory although the idea might be traced back earlier. In 1999, Seiberg and Witten proposed~\cite{SW} that some low-energy effective theory of open strings with a nontrivial background can be described by a noncommutative filed theory.  Subsequently, the great progress on noncommutative field theory has been made, see, for instance, the review articles~\cite{Review}.

Black holes, originated from Einstein's field equations of general relativity, 
have played an important role in quantum gravity, and the relevant thermodynamics has acquired great strides~\cite{RW,SC}. In particular, the introduction of a negative cosmological constant makes black holes present rich thermodynamic behaviors. The variation of the cosmological constant $\Lambda$ in the first law of black hole thermodynamics has been widely accepted. Crucially, the cosmological constant can be interpreted as the thermodynamic pressure $P$ with
\begin{equation}
P=-\frac{\Lambda}{8\pi}=\frac{(n-1)(n-2)}{16 l^2 \pi},\label{pres}
\end{equation}
where $n$ stands for the dimension of spaetime and $l$ represents the curvature radius of the AdS spacetime. When the cosmological constant corresponds to the pressure as a thermodynamic variable, the black hole mass $M$ can be identified with the enthalpy, rather than the internal energy. Then, the conjugate variable corresponding to the cosmological constant is {\em thermodynamic volume} with $V=(\partial M/\partial P)_{S}$. In this way, one can describe the black hole thermodynamic behaviors in an {\em extended phase space} with the pressure and volume as thermodynamic variables. In particular, by an analogy with the van der Waals fluid,  the equation of state of a charged AdS black hole has attained increasing interest~\cite{BPD,CEJM,KM}.

In 1993, Susskind suggested \cite{LS} that stringy effects cannot be neglected in the string/black hole correspondence principle. Now it is known that noncommutative geometry inspired black holes~\cite{NSS} (sometimes in short {\em noncommutative black holes}) contain stringy effects, where such an effect is similar in some sense to that of noncommutative field theory induced by string theory. One way to introduce noncommutative (stringy) effects into black holes, as suggested in ref.~\cite{NSS}, is to modify energy-momentum tensors in terms of smeared matter distributions. Specifically, the point-like $\delta$-function mass density is replaced by the Gaussian smeared matter distribution in the right hand side of Einstein's field equations, while no changes are made in the left hand side. In this way, a self-regular black hole solution with noncummutative effects but without curvature singularities is given. Since the work of ref.~\cite{NSS},  a lot of developments have been made, such as generalizations to high-dimensional black holes~\cite{TG}, charged black holes~\cite{ANSS}, high-dimensional charged black holes~\cite{SSN}, and to the topics in other diverse aspects~\cite{PN,SS,MXZ}.


Besides the Gaussian smeared matter distribution, non-Gaussian smeared matter distributions have also been considered, such as the Lorentzian smeared mass distribution \cite{NM}, the Rayleigh distribution \cite{MY}, and the ring-type distribution~\cite{Park}, etc. 
In fact, the Gaussian smeared matter distribution is not always required~\cite{POS}, for instance, the ring-type smeared matter distribution in $(2+1)$-dimensional spacetime has been found~\cite{Park} to have quite interesting features in phase transition and soliton-like behaviors of black holes.

Naturally, in order to acquire more understanding to the noncommutative high-dimensional Schwarzschild-Tangherlini anti-de Sitter black hole, we apply the non-Gaussian smeared matter distribution proposed in ref.~\cite{Park} to the (ordinary) high-dimensional Schwarzschild-Tangherlini anti-de Sitter black hole 
and study the thermodynamic behaviors, in particular the phase transition, of such a noncommutative black hole. We find that the Maxwell equal area law maintains  for this noncommutative AdS black hole if the Hawking temperature stays in a specific range. As a byproduct, we indicate that the Gaussian smeared matter distribution is not applicable for the 6- and higher-dimensional black holes in accordance with the so-called hoop conjecture\footnote{The matter mean radius of a black hole related to some mass distribution should not be larger than the horizon radius of the relevant extreme black hole in order to ensure the formation of a black hole.}~\cite{HC}.

The arrangement of this paper is as follows.
In section \ref{sec2}, we give an introduction of the noncommutative high-dimensional Schwarzschild-Tangherlini AdS black hole that is  associated with the non-Gaussian smeared matter distribution proposed in ref.~\cite{Park}. Then, the thermodynamic quantities of the noncommutative black hole are calculated and the characteristics of phase transitions are analyzed in detail in section \ref{sec3}. Finally, a brief summary is made in section \ref{sec4}.

\section{Noncommutative high-dimensional AdS black hole}\label{sec2}
We consider a high-dimensional ($n \geq 4$), neutral, and non-rotating Schwarzschild-Tangherlini anti-de Sitter black hole \cite{ST} with a negative cosmological constant. The metric reads as
\begin{equation}
\text{ds}^2=-f(r)dt^2+\frac{dr^2}{f(r)}+r^2 d{\Omega}^{2}_{n-2},
\end{equation}
where $d\Omega_{n-2}^2$ is the square of line element on an $(n-2)$-dimensional unit sphere and the general form of function $f(r)$ takes the form,\footnote{The geometric units, $\hbar=c=k_{_B}=G=1$, are adopted throughout this paper.}
\begin{equation}
f(r)=1-\frac{16\pi m(r)}{(n-2)\omega r^{n-3}}+\frac{r^2}{l^2},\label{xianyuan}
\end{equation}
where $\omega$ denotes the area of  an $(n-1)$-dimensional unit sphere\footnote{$\omega=\frac{2\pi^{\frac{n-1}{2}}}{\Gamma\left(\frac{n-1}{2}\right)}$, where $\Gamma(x)$ is the gamma function.} and $m(r)$ stands for the black hole mass distribution we shall choose.


In this $n$-dimensional AdS spacetime we adopt the non-Gaussian mass density of black holes proposed by ref.~\cite{Park}, \begin{equation}
\rho(r)=A r^k e^{-\left(\frac{r}{2\sqrt{\theta}}\right)^2},\label{fenbu}
\end{equation}
where $\sqrt{\theta}$ is the noncommutative parameter with the dimension of length, $k$ is a non-negative integer,
$k=0,1,2,\cdots$, and $A$ is a normalization constant that can be fixed by using the constraint: $\int_0^{\infty}\rho(r) dV_{n-1}=M$,
\begin{equation}
A=\frac{M}{\pi^{\frac{n-1}{2}}\left(2\sqrt{\theta}\right)^{n+k-1}} \,\frac{\Gamma\left(\frac{n-1}{2}\right)}{\Gamma\left(\frac{n+k-1}{2}\right)},\label{paraA}
\end{equation}
where the parameter $M$ is the ADM mass of black holes, and $dV_{n-1}$ is an $(n-1)$-dimensional volume element.
We note that this kind of non-Gaussian smeared mass densities is general, which means that it includes the Gaussian distribution of $k=0$ and the Rayleigh distribution of $k=1$ as special cases.

Now the corresponding mass distribution can be derived from eqs.~(\ref{fenbu}) and (\ref{paraA}),
\begin{equation}
m(r)=\int_0^{r}\rho(r^{\prime}) dV_{n-1}=\frac{M}{\Gamma\left(\frac{n+k-1}{2}\right)}\gamma\left(\frac{n+k-1}{2},\left(\frac{r}{2\sqrt{\theta}}\right)^2\right),\label{masfenbu}
\end{equation}
where $\gamma(a,x)$ is the lower incomplete gamma function.


Moreover, using eqs.~(\ref{xianyuan}) and (\ref{masfenbu}) we can get the ADM mass $M$ in terms of the black hole horizon radius $r_h$,
\begin{equation}
M=\frac{(n-2)\omega \Gamma\left(\frac{n+k-1}{2}\right)}{16\pi \gamma\left(\frac{n+k-1}{2},\left(\frac{r_h}{2\sqrt{\theta}}\right)^2\right)}\left(r_h^{n-3}+\frac{r_h^{n-1}}{l^2}\right), \label{mas}
\end{equation}
where $r_h$ is thought to be the largest real root of $f(r)=0$.
When taking the commutative limit $\theta\rightarrow 0$,\footnote{In the noncommutative case, the noncommutative effect can be neglected when the horizon radius $r_h$ is becoming large. Only in this sense, the commutative limit is equivalent to the large horizon radius limit.}
we can see that the black hole mass turns out to be the known one \cite{ST},
\begin{eqnarray}
M \rightarrow\frac{(n-2)\omega}{16\pi}\left(r_h^{n-3}+\frac{r_h^{n-1}}{l^2}\right),\label{mas10}
\end{eqnarray}
which shows the consistency of our noncommutative generalization.

For the sake of convenience, we introduce two dimensionless parameters $b$ and $x_h$ defined by
\begin{equation}
b:=\frac{2\sqrt{\theta}}{l}, \qquad x_h:=\frac{r_h}{2\sqrt{\theta}}, \label{daihuan1}
\end{equation}
and rewrite eq.~(\ref{mas}) as follows,
\begin{equation}
\frac{M}{\left(2\sqrt{\theta}\right)^{n-3}}=\frac{(n-2)\omega \Gamma\left(\frac{n+k-1}{2}\right)}{16\pi \gamma\left(\frac{n+k-1}{2},x_h^2\right)}\left(x_h^{n-3}+b^2 x_h^{n-1}\right). \label{mas11}
\end{equation}
The purpose is to express the relation of a black hole mass with respect to a horizon radius in terms of the single parameter $b$. For a noncommutative spacetime with a small but finite value of $\sqrt{\theta}$, this parameter $b$ is becoming small when the curvature radius of the AdS spacetime $l$ is large, which corresponds to an asymptotic Minkowski spacetime; but $b$ is becoming large when $l$ is small, which corresponds to a spacetime with a strong AdS background. As a result, the range of parameter $b$ is usually from zero to infinity.  Because $2\sqrt{\theta}$ can be regarded as the minimal length of the related noncommutative spacetime, $\frac{M}{\left(2\sqrt{\theta}\right)^{n-3}}$ can be understood as the black hole mass in the unit of the Planck mass and $\frac{r_h}{2\sqrt{\theta}}$ as the horizon radius in the unit of the Planck length if the minimal length is dealt with as the order of the Planck length. 


The horizon radius of extreme black holes $r_0$ satisfies the relation $\partial M /\partial r_h =0$ that can be written with the help of  eq.~(\ref{mas}) or eq.~(\ref{mas11}) as follows:
\begin{equation}
G(n,k;x_0)=n-1-\frac{2}{1+b^2 x_0^2},\label{extreradi}
\end{equation}
where the parameter $x_0$ is defined by
\begin{equation}
x_0 :=\frac{r_0}{2\sqrt{\theta}} \label{daihuan2},
\end{equation}
whose meaning is a horizon radius of extreme black holes in the unit of the minimal length, and the function $G(n,k;x)$ is defined by
\begin{equation}
G(n,k;x):=\frac{2 x^{n+k-1} e^{-x^2}}{\gamma\left(\frac{n+k-1}{2},
x^2\right)},\label{tezheng}
\end{equation}
where $x:=\frac{r}{2\sqrt{\theta}}$. See Appendix for a detailed analysis of the function $G(n,k;x)$.

Before solving eq.~(\ref{extreradi}), we have to consider the {\em hoop conjecture} condition~\cite{HC}, that is, the matter mean radius  of the mass distribution $\bar{r}$ should not be larger than the horizon radius of extreme black holes. The matter mean radius that relates to the non-Gaussian mass density distribution (eq.~(\ref{fenbu})) can be calculated,
\begin{equation}
\bar{r}=\int_0^{\infty}r\, \frac{\rho(r)}{M} dV_{n-1}=2\sqrt{\theta}\, \frac{\Gamma(\frac{n+k}{2})}{\Gamma(\frac{n+k-1}{2})}.
\end{equation}
That is,  the {\em hoop conjecture} implies $\bar{r} \leq r_0$, or $\bar{x} \leq x_0$, where $\bar{x} :=\frac{\bar{r}}{2\sqrt{\theta}}$. If not, the black hole could not be formed.

Considering the constraint $ 0 < b < \infty$ and the characteristics of the function $G(n,k;x)$ that are listed in Appendix, we obtain from eq.~(\ref{extreradi}) the range of the horizon radius of extreme black holes, $x_{*} < x_0 < \tilde{x}$, where $x_{*}$ is the root of the equation $G(n,k;x_{*})=n-1$  and $\tilde{x}$ is the root of the equation $G(n,k;\tilde{x})=n-3$.
Further due to the {\em hoop conjecture}, the range of $x_0$ reads as
\begin{equation}
\text{Max}\{\bar{x},x_{*}\}< x_0 < \tilde{x}. \label{range}
\end{equation}
Then, using eqs.~(\ref{extreradi})-(\ref{range}) and keeping in mind that $n$ and $k$ are integers, 
we can get the  allowed values of $k$ at various dimensions\footnote{In general, the dimension $n$ can take any positive integers. However, we prefer to consider the range of $n$ from four to eleven in physics.} in the following two cases.

(i) When $\bar{x} > x_{*}$, we can see that $b$ is further constrained to the small range, $b \in \left(0, b|_{x_0=\bar{x}}\right)$, and obtain the results given in Table \ref{biao1}.

\begin{table}[!hbp]
\begin{center}
\begin{tabular}{|c|*{9}{|c}}
\hline
\multicolumn{9}{|c|} {For various dimensions $n$, the allowed values of $k$} \\ \hline
n & $4$ & $5$ & $6$ & $7$ & $8$ & $9$ & $10$ & $11$ \\ \hline
k & $[0,5]$ & $[0,9]$ & $[4,15]$ & $[8,23]$ & $[14,33]$ & $[22,44]$ & $[32,56]$ & $[43,70]$\\ \hline
\end{tabular}
\end{center}
\caption{When $\bar{x} > x_{*}$, according to eqs.~(\ref{extreradi})-(\ref{range}), we list the allowed values of $k$ for various $n$,     where $b$ is further constrained to the small range, $b \in \left(0, b|_{x_0=\bar{x}}\right)$.}
\label{biao1}
\end{table}

(ii) When $\bar{x} < x_{*}$, we find that $b$ has no extra constraints, $b \in (0, \infty)$, and obtain the results given in Table \ref{biao2}.

\begin{table}[!hbp]
  \begin{center}
\begin{tabular}{|c|*{9}{|c}}
\hline
\multicolumn{9}{|c|} {For various dimensions $n$, the allowed values of $k$} \\ \hline
n & $4$ & $5$ & $6$ & $7$ & $8$ & $9$ & $10$ & $11$ \\ \hline
k & $\geq 6$ & $\geq 10$ & $\geq 16$ & $\geq 24$ & $\geq 34$ & $\geq 45$ & $\geq 57$ & $\geq 71$\\ \hline
\end{tabular}
  \end{center}
  \caption{When $\bar{x} < x_{*}$, according to eqs.~(\ref{extreradi})-(\ref{range}), we list the allowed values of $k$ for various $n$, where $b$ has no extra constraints, $b \in (0, \infty)$.}
\label{biao2}
\end{table}

It is remarkable that for the high-dimensional Schwarzschild-Tangherlini AdS black hole the Gaussian smeared mass distribution (the case $k=0$) is not applicable to the $d \geq 6$ dimensions.

Now we turn to the extremal horizon radius described by eqs.~(\ref{extreradi})-(\ref{tezheng}). The numerical results\footnote{In all of the Tables and Figures of this paper, the values of the dimensionless parameter $b$ we set coincide with the {\em hoop conjecture}. That is to say, these values satisfy eqs.~(\ref{extreradi})-(\ref{range}).} are shown in Table \ref{biao3}. We can see that the extremal horizon radius increases when the power $k$ increases for a fixed $n$, which is obvious because the matter mean radius increases. On the contrary, the extremal horizon radius decreases when the dimension $n$ increases for a fixed $k$, indicating that the higher the dimension is, the smaller the extremal horizon radius is. 

\begin{table}[!hbp]
\begin{center}
\begin{tabular}{|c|*{9}{|c}}
\hline
\multicolumn{9}{|c|} {The extremal horizon radius $x_0=r_0/(2\sqrt{\theta})$} \\ \hline
\backslashbox{k}{n}
& $4$ & $5$ & $6$ & $7$ & $8$ & $9$ &$10$ &$11$ \\ \hline \hline
0 & $1.5059$ & $1.3361$ & $-$ & $-$ & $-$ & $-$ & $-$ & $-$\\ \hline
1 & $1.7860$ & $1.6099$ & $-$ & $-$ & $-$ & $-$ & $-$ & $-$  \\ \hline
3 & $2.2097$ & $2.0299$ & $-$ & $-$ & $-$ & $-$ & $-$ & $-$ \\ \hline
4 & $2.3838$ & $2.2037$ & $2.1056$ & $-$ & $-$ & $-$ & $-$ & $-$ \\ \hline
5 & $2.5419$ & $2.3619$ & $2.2626$ & $-$ & $-$ & $-$ & $-$ & $-$ \\ \hline
8 & $2.9502$ & $2.7716$ & $2.6705$ & $2.6027$ & $-$ & $-$ & $-$ & $-$\\ \hline
10 & $3.1849$ & $3.0076$ & $2.9059$ & $2.8371$ & $-$ & $-$ & $-$ & $-$\\ \hline
15 & $3.6899$ & $3.5159$ & $3.4139$ & $3.3437$ & $3.2916$ & $-$ & $-$ & $-$\\ \hline
18 & $3.9542$ & $3.7821$ & $3.6803$ & $3.6097$ & $3.5568$ & $-$ & $-$ & $-$ \\ \hline
19 & $4.0375$ & $3.8660$ & $3.7642$ & $3.6935$ & $3.6405$ & $-$ & $-$ & $-$ \\ \hline
20 & $4.1186$ & $3.9477$ & $3.8461$ & $3.7753$ & $3.7221$ & $-$ & $-$ & $-$ \\ \hline
25 & $4.4973$ & $4.3292$ & $4.2281$ & $4.1571$ & $4.1034$ & $4.0608$ & $-$ & $-$ \\ \hline
35 & $5.1550$ & $4.9919$ & $4.8921$ & $4.8212$ & $4.7669$ & $4.7235$ & $4.6878$ & $-$ \\ \hline
39 & $5.3911$ & $5.2298$ & $5.1305$ & $5.0597$ & $5.0054$ & $4.9618$ & $4.9258$ & $-$ \\ \hline
40 & $5.4482$ & $5.2873$ & $5.1882$ & $5.1174$ & $5.0630$ & $5.0194$ & $4.9834$ & $-$ \\ \hline
45 & $5.7236$ & $5.5647$ & $5.4662$ & $5.3956$ & $5.3412$ & $5.2975$ & $5.2612$ & $5.2305$ \\ \hline
50 & $5.9839$ & $5.8268$ & $5.7291$ & $5.6587$ & $5.6043$ & $5.5604$ & $5.5240$ & $5.4931$ \\ \hline
\end{tabular}
\end{center}
\caption{The numerical results of the extremal horizon radius $x_0$ for different dimensions $n$ and different powers $k$ are  listed, where $b=0.0447$. A hyphen means that the corresponding black hole is forbidden by the hoop conjecture, so no extremal horizon radius exists.}
\label{biao3}
\end{table}

\begin{figure}
\centering
\subfloat[$n=5, b=0.0447$, and $k=0,1,3,5$, respectively, from left to right.]{\includegraphics[width=125mm]{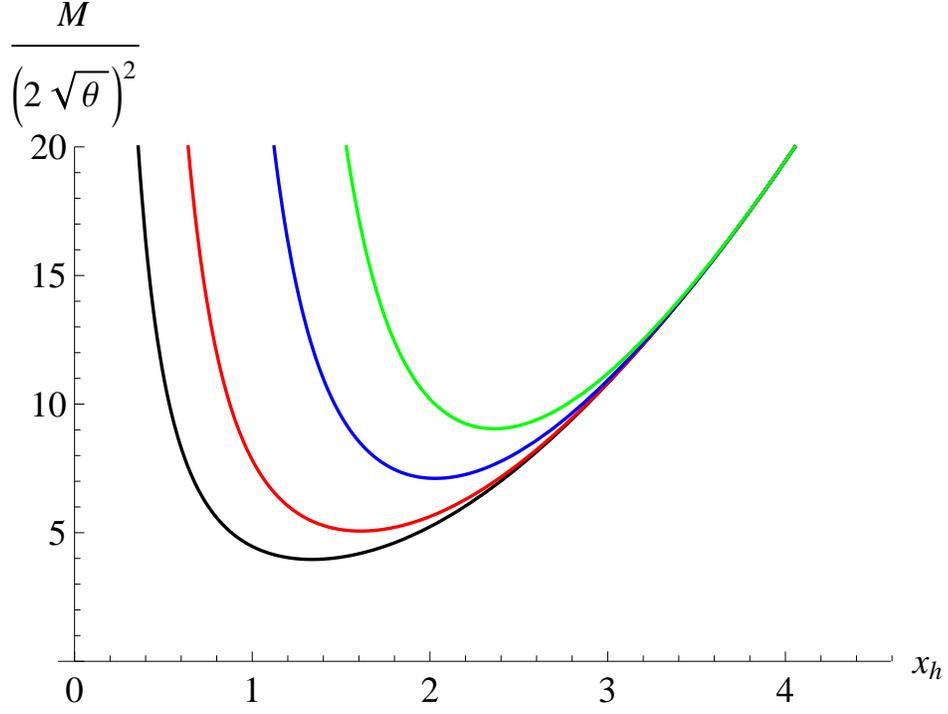}}\\
\subfloat[$n=7, k=25$, and $b=0.0632, 0.0994, 0.134406, 0.1499, 0.1721$, respectively, from bottom to top, where the orange dashed curve corersponds to the critical value of $b$ at which the maximum and minimum Hawking temperatures just disappear. See also Figure \ref{tu2} for this critical case.]{\includegraphics[width=125mm]{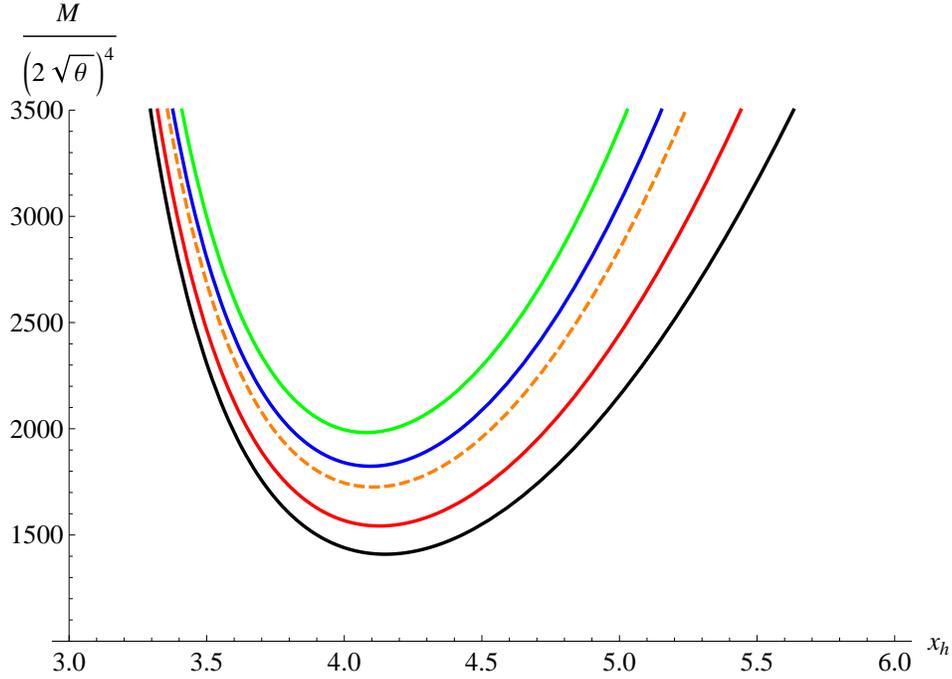}}
\caption{Plots of the relations of $M$ with respect to $x_h$.}
\label{tu1}
\end{figure}

\begin{figure}
\begin{center}
\includegraphics[width=125mm]{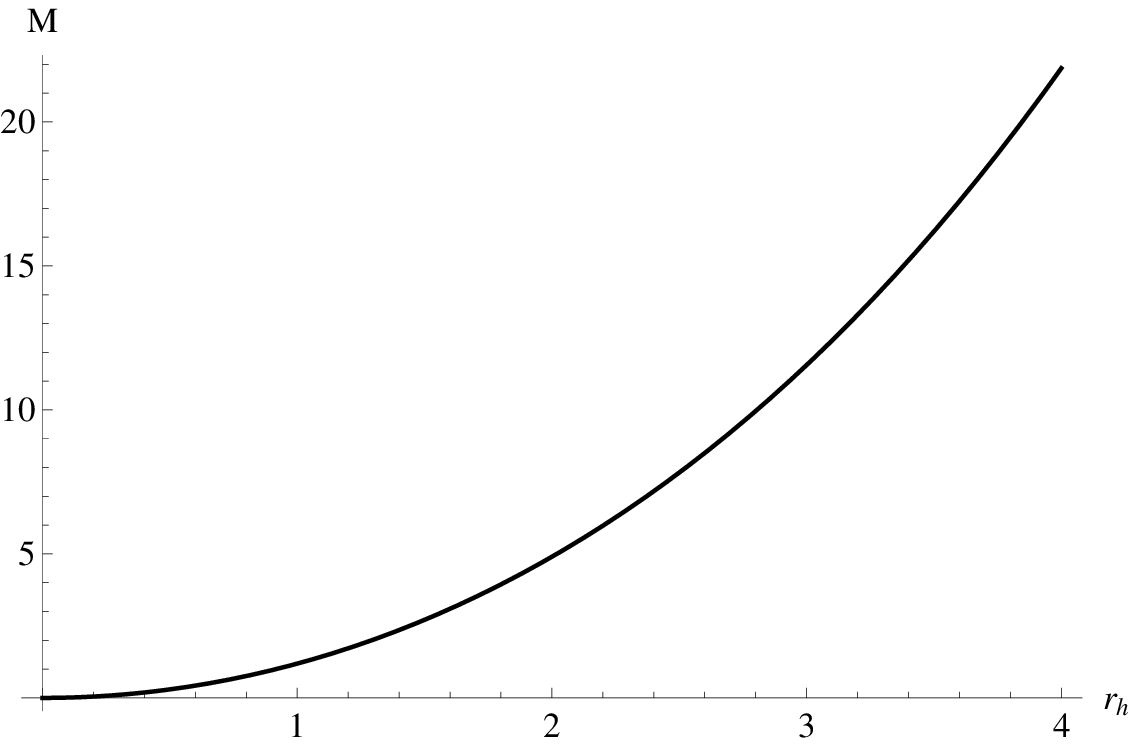}
\end{center}
\caption{Plot of the relation of $M$ with respect to $r_h$ for the commutative limit  $\theta\rightarrow 0$,  see eq.~(\ref{mas10}),  where $l=10$ and $n=5$.}
\label{tu11}
\end{figure}

For analyzing the relation of the black hole mass with respect to the horizon radius, we adopt the numerical method because eq.~(\ref{mas}) or eq.~(\ref{mas11}) cannot be solved analytically. For instance, taking $n=5$ and setting different powers, say, $k=0,1,3,5$, we plot the function $M=M(x_h)$ in Figure \ref{tu1}. One sees that there is one minimum mass $M_0$ when the horizon radius takes the extremal horizon radius $x_0$. 
That is to say, the extreme black hole exists. It is worth illustrating three cases: (i) when $M > M_0$, there exists a black hole with two horizons; (ii) when $M = M_0$, there exists an extreme black hole; (iii) when $M < M_0$, no black hole exists. In addition, under the commutative limit $\theta\rightarrow 0$ or the large horizon radius $r_h$ limit, the extreme black hole disappears and the noncommutative black hole turns back to the commutative one, see Figure \ref{tu11}.

\section{Thermodynamic analysis}\label{sec3}
In this section we analyze the phase transition of the noncommutative black hole introduced in the above section, and also investigate the relevant thermodynamic features, such as the entropy, Gibbs free energy, and equation of state.

\subsection{Phase transition}
If a phase transition exists, the noncommutative black hole will attain a stable or an  equilibrium configuration. 
When a phase transition happens, some critical phenomena will appear. For example, the heat capacity at a constant pressure will be divergent or the temperature will approach an extremum. In the following we analyze the phase transition through studying the divergence of the heat capacity.

The heat capacity at a constant pressure is defined by
\begin{equation}
C_p :=\left(\frac{\partial M}{\partial T_h}\right)_p=\frac{\partial M}{\partial r_h}\left(\frac{\partial T_h}{\partial r_h}\right)^{-1}.\label{cap}
\end{equation}
Considering the Hawking temperature $T_h=\frac{f^{\prime}(r_h)}{4\pi}$ and using eqs.~(\ref{xianyuan}), (\ref{masfenbu}), and  (\ref{daihuan1}), we obtain
\begin{equation}
2\sqrt{\theta}T_h
=\frac{1}{4\pi}\left\{\frac{n-3-G(n,k;x_h)}{x_h}+b^2 x_h\left[n-1-G(n,k;x_h)\right]\right\}.\label{tem11}
\end{equation}
Again using eqs.~(\ref{mas}) and (\ref{tem11}) we derive the factors of the numerator and denominator of eq.~(\ref{cap}) multiplied by suitable normalization factors, respectively,
\begin{equation}
\begin{aligned}
\left(\frac{1}{2\sqrt{\theta}}\right)^{n-4}\frac{\partial M}{\partial r_h}=&\frac{(n-2)\omega \Gamma\left(\frac{n+k-1}{2}\right)x_h^{n-2}}{16\pi \gamma\left(\frac{n+k-1}{2},x_h^2\right)}\left\{b^2 \left[n-1-G(n,k;x_h)\right]+\frac{n-3-G(n,k;x_h)}{x_h^2}\right\},\\
\vspace{0.4cm}
\left(\frac{1}{2\sqrt{\theta}}\right)^{-2}\frac{\partial T_h}{\partial r_h}=& \frac{1}{4\pi}\left\{b^2 \left[n-1-G(n,k;x_h)\right]-\frac{n-3-G(n,k;x_h)}{x_h^2}\right.\\
&\,\,\,\,\,\,\,\,\,\,\,\, \left. -\left(b^2 x_h +\frac{1}{x_h}\right)G^{\prime}(n,k;x_h)\right\},\label{fenzifenmu}
\end{aligned}
\end{equation}
where $G^{\prime}(n,k;x_h)$ stands for the first order derivative of $G(n,k;x_h)$ with respect to $x_h$.

For the black hole with a large horizon radius $r_h$ or under the limit $\theta\rightarrow 0$, the Hawking temperature $T_h$ (eq.~(\ref{tem11})) turns back to that of the commutative black hole~\cite{BC},
\begin{eqnarray}
T_h \rightarrow\frac{1}{4\pi}\left[\frac{n-3}{r_h}+\frac{(n-1)r_h}{l^2}\right].\label{temp}
\end{eqnarray}
In addition, using eqs.~(\ref{cap})-(\ref{fenzifenmu}) we observe that the heat capacity tends to the commutative formulation~\cite{RD} under $\theta\rightarrow 0$ or the large horizon radius $r_h$ limit,
\begin{eqnarray}
C_p \rightarrow \frac{(n-2)\omega}{4} \frac{(n-3)l^2 r_h^{n-2}+(n-1)r_h^{n}}{(n-1)r_h^2-(n-3)l^2}.\label{cap1}
\end{eqnarray}
The above limits show that the noncommutative generalizations of the Hawking temperature and the heat capacity are reasonable.

\begin{figure}
\centering
\subfloat[$n=5$, $b=0.0447$, and $k=0,1,3,5$, respectively, from left to right. For $k=3$ ({\color{blue}blue} curve), we give the critical radii and their corresponding extremal temperatures, i.e. $x_c=3.0185$, $2\sqrt{\theta}T_{max}=0.05053$ and $x_c|_{{x_h}\uparrow}=15.8189$, $2\sqrt{\theta}T_{min}=0.02012$.]{\includegraphics[width=125mm]{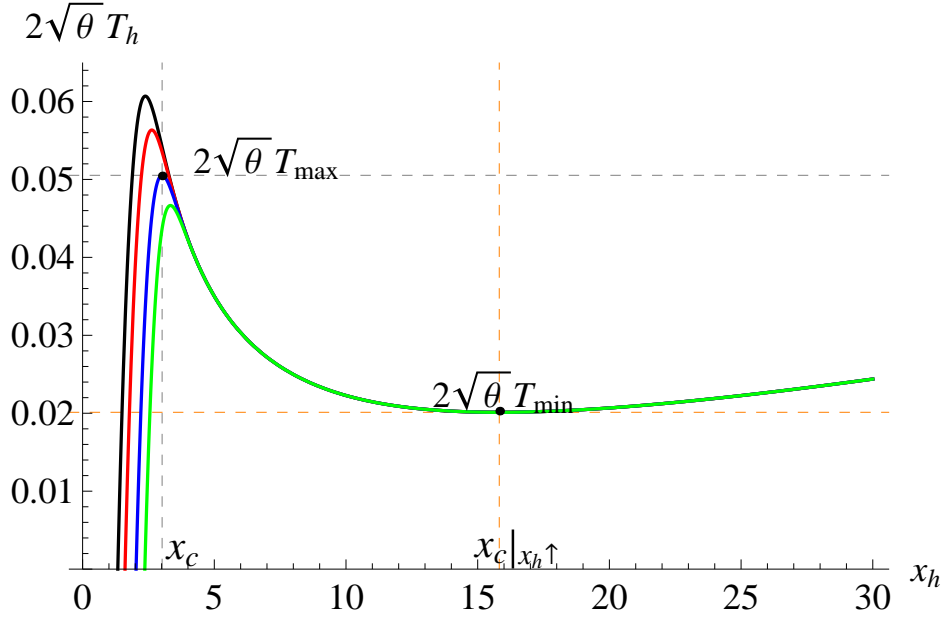}}\\
\subfloat[$n=7, k=25$, and $b=0.0932, 0.1194, 0.134406, 0.1471, 0.1551$, respectively, from bottom to top. The orange dashed curve is the critical curve at $b_c=0.134406$, below which the maximum and minimum temperatures exist but above which no such temperatures exist.]{\includegraphics[width=125mm]{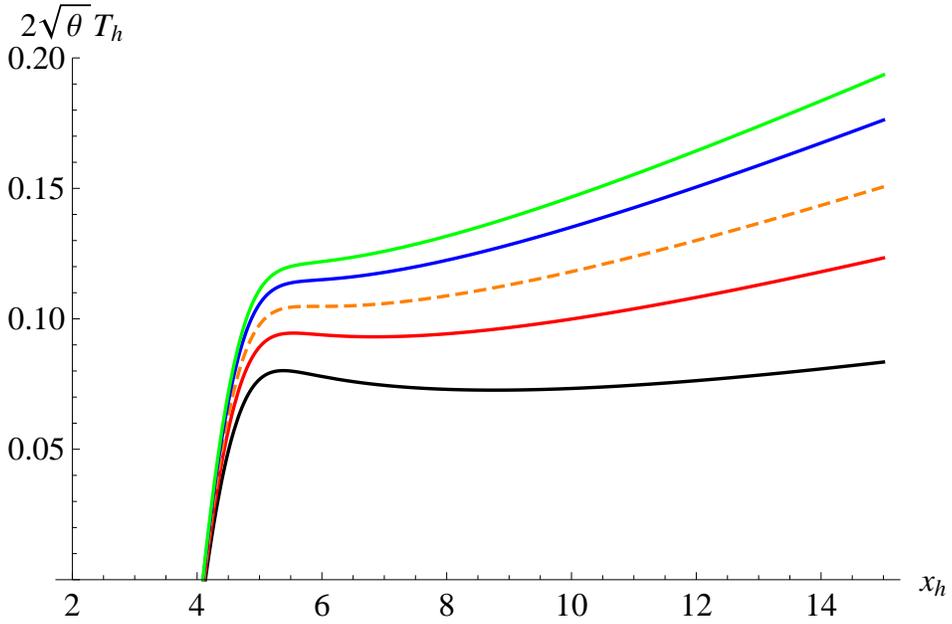}}
\caption{Plots of the relations of $T_h$ with respect to $x_h$.}
\label{tu2}
\end{figure}

\begin{figure}
\begin{center}
\includegraphics[width=125mm]{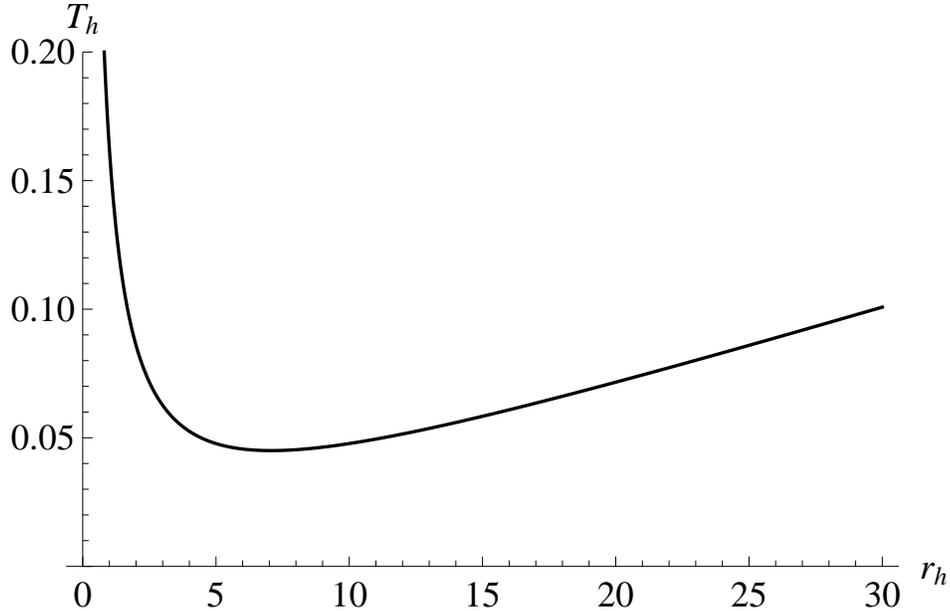}
\end{center}
\caption{Plot of the relation of $T_h$ with respect to $r_h$ for the commutative limit  $\theta\rightarrow 0$, see eq.~(\ref{temp}), where $l=10$ and $n=5$.}
\label{tu22}
\end{figure}

Eq.~(\ref{tem11}) is plotted in Figure \ref{tu2}. For the extreme black hole, the temperature vanishes at the extremal horizon radius. For the non-extreme black holes, there are maximum temperature $2\sqrt{\theta}T_{max}$ and minimum temperature $2\sqrt{\theta}T_{min}$ at critical points labeled by $x_c$ and $x_c|_{{x_h}\uparrow}$, respectively, where $x_c|_{{x_h}\uparrow}$ means the critical radius under the large $x_h$ limit. In the second diagram of Figure \ref{tu2}, we observe that the maximum and minimum temperatures are gradually disappearing when the parameter $b$ is approaching the critical value $b_c=0.134406$. This feature happens for the noncommutative black hole with a strong noncommutativity with respect to the AdS radius, which can be seen clearly from the definition of the parameter in eq.~(\ref{daihuan1}).
While for the commutative situation, that is, the case under $\theta\rightarrow 0$, there is only one  minimum temperature shown in Figure \ref{tu22}.

\begin{figure}
\centering
\subfloat[$n=5, b=0.0447$, and $k=0,1,3,5$, respectively, from left to right.]{\includegraphics[width=125mm]{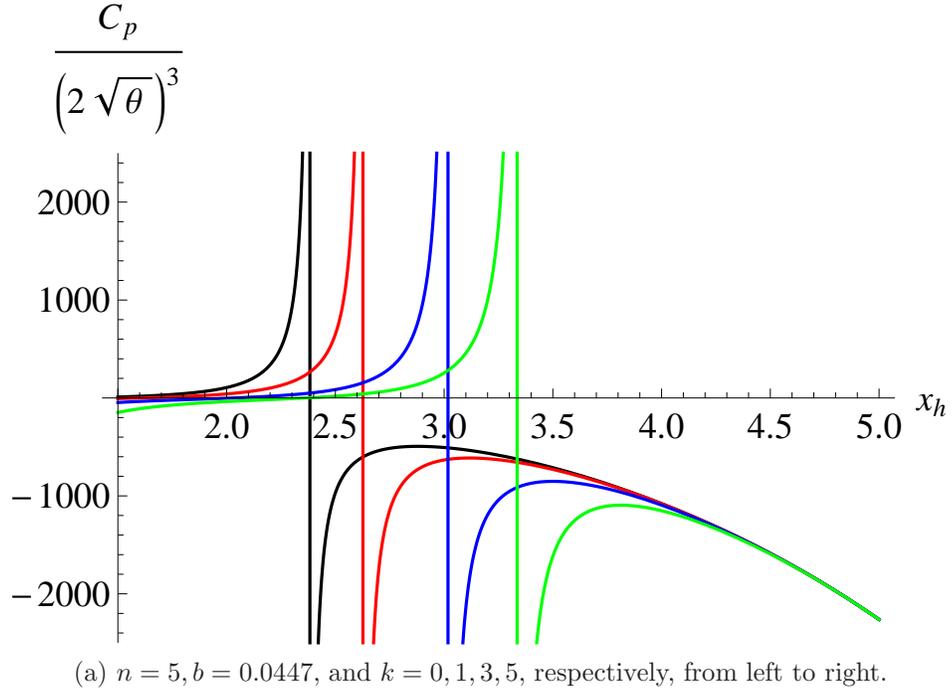}}\\
\vspace{4mm}
\subfloat[$n=7$, $k=25$, and $b=0.1222$ (\text{{\color{black}black}}), 0.1295 (\text{{\color{blue}blue}}), 0.134406 (\text{\textcolor[rgb]{1.00,0.50,0.00}{orange}}), 0.1360 (\text{{\color{green}green}}), 0.1395 (\text{{\color{yellow}yellow}}), respectively. The orange dashed curves are the critical curves at $b_c=0.134406$.]{\includegraphics[width=125mm]{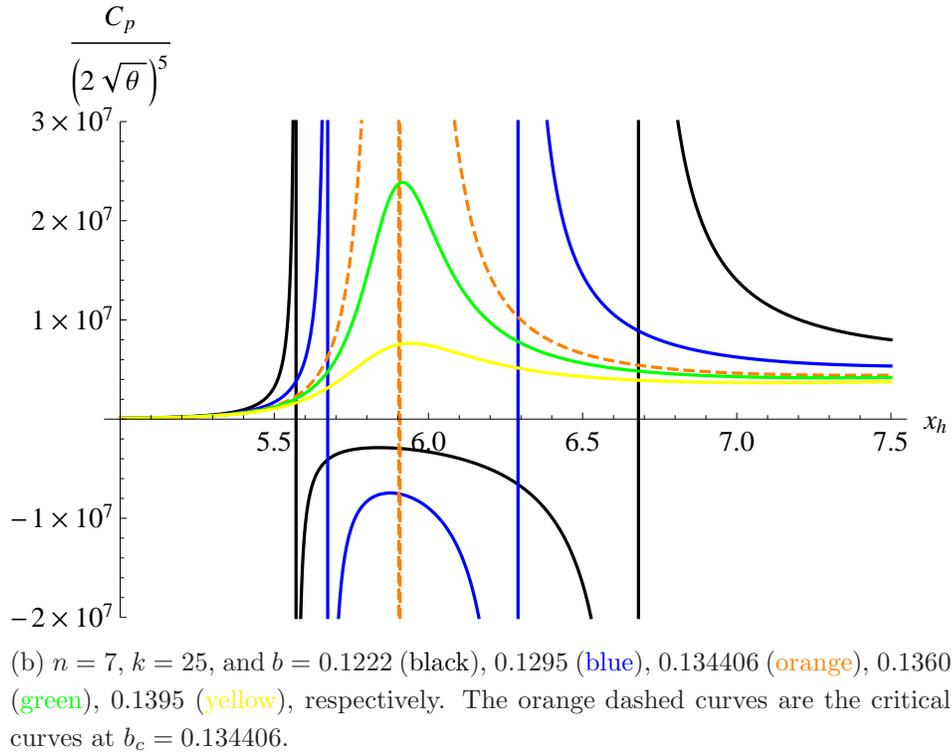}}
\caption{Plots of the relations of $C_p$ with respect to $x_h$.}
\label{tu3}
\end{figure}

We know that the thermodynamic stability of black holes is determined by the heat capacity. The black hole is locally stable for $C_p >0$, but unstable for $C_p <0$. It is shown in Figure \ref{tu3} that the heat capacity is divergent at the critical radius $x_c$ or $x_c|_{{x_h}\uparrow}$, which implies that the  phase transition occurs at $x_c$ or $x_c|_{{x_h}\uparrow}$.
When the parameter $b$ increases, $x_c$ and $x_c|_{{x_h}\uparrow}$ are approaching to each other, and finally the divergence will disappear.  This gives rise to $C_p >0$, which means that the black hole is locally stable, see the green and yellow lines in the second diagram of Figure \ref{tu3}. In other words, no phase transition happens if $b \geq b_c$. Incidentally, the Gaussian matter distribution in the 5-dimensional AdS spacetime, i.e. the case of $n=5$ and $k=0$, was investigated in ref.~\cite{sss}, which can be dealt with as a specific case of our results (the black curves in the first diagram of Figure \ref{tu3}).

\begin{figure}
\begin{center}
\includegraphics[width=125mm]{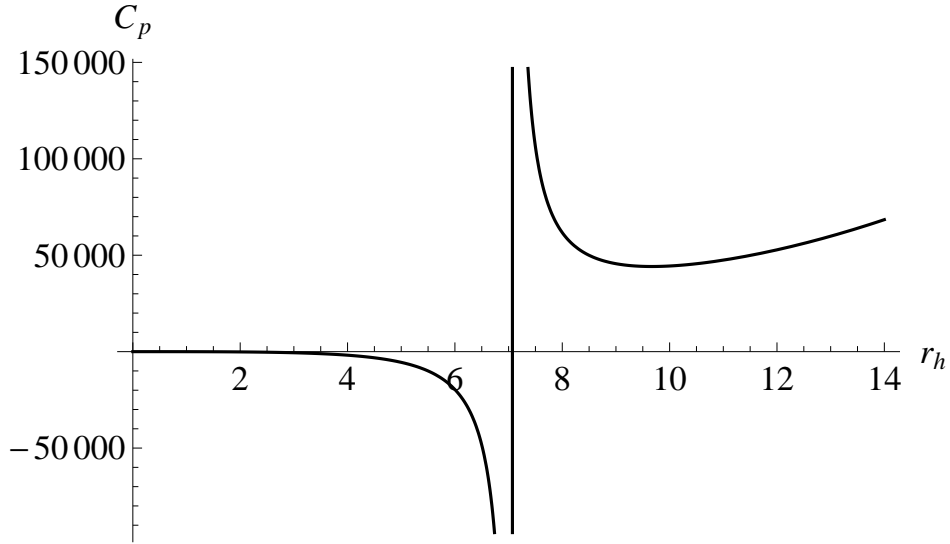}
\end{center}
\caption{Plot of the relation of $C_p$ with respect to $r_h$ for the commutative limit $\theta\rightarrow 0$, see eq.~(\ref{cap1}),  where $l=10$ and $n=5$.}
\label{tu33}
\end{figure}

The critical radius $x_c=r_c/(2\sqrt{\theta})$ can be obtained by setting the denominator of eq.~(\ref{cap}) equal  zero,
\begin{equation}
\left.\frac{\partial T_h}{\partial r_h}\right|_{r_h=r_c}=0. \label{cap0}
\end{equation}
Although the above equation cannot be solved analytically,
we can obtain its asymptotic behavior under the commutative limit $\theta\rightarrow 0$ or the large horizon radius limit,
\begin{equation}
r_c \rightarrow  \left({\frac{n-3}{n-1}}\right)^{1/2}l. \label{capra}
\end{equation}
Instead of analytical results of eq.~(\ref{cap0}),  we list some numerical results in Table \ref{biao4} for a certain value of the parameter $b$. Note that eq.~(\ref{cap0}) has two roots for a fixed dimension $n$: one root is small, given in Table \ref{biao4} for different $k$, which corresponds to the noncommutative case, and the other root is big, listed in the last line, which corresponds to the large horizon radius limit. From the data of the table we know that the first phase transition occurs at a small critical horizon radius $x_c$, where the black hole becomes locally unstable from locally stable, and second phase transition occurs at a large critical horizon radius, i.e., at $x_c|_{{x_h}{\uparrow}}$, where the black hole becomes locally stable from locally unstable. Incidentally, for the commutative black hole, only one phase transition happens at the position given by eq.~(\ref{capra}), see Figure \ref{tu33}.

We can see from Table \ref{biao4} that the critical radius $x_c$ at which the first phase transition occurs increases when the power $k$ increases for a fixed $n$, which is quite natural because the extremal horizon radius $x_0$ increases (cf. Table \ref{biao3}). However, the situation is complicated when the dimension $n$ increases for a fixed $k$. For a small $k$ of zero to three, the critical radius $x_c$ decreases when the dimension $n$ increases. While for a large $k$ equal to and larger than four, an anomaly appears. That is, the critical radius $x_c$ decreases at the beginning and increases later. For instance, when $k=8$, the critical radius $x_c$ decreases when $n$ takes four to six, but it suddenly increases when $n$ takes seven. This feature shows that an anomalous trend of critical radii exists in the first phase transition for the 6- and higher-dimensional black holes, see Table \ref{biao4} for the details. Incidentally, no such an anomaly exists in the second phase transition, where the corresponding critical radius $x_c|_{{x_h}\uparrow}$ increases when the dimension $n$ increases for a fixed $k$,  see the last line of Table \ref{biao4} for the details.

\begin{table}[!hbp]
\begin{center}
\begin{tabular}{|c|*{9}{|c}}
\hline
\multicolumn{9}{|c|} {The critical horizon radius $x_c=r_c/(2\sqrt{\theta})$ at which a phase transition happens} \\ \hline
\backslashbox{k}{n}
& $4$ & $5$ & $6$ & $7$ & $8$ & $9$ & $10$ & $11$\\ \hline \hline
0 & $2.3991$ & $2.3826$ & $-$ & $-$ & $-$ & $-$ & $-$ & $-$ \\ \hline
1 & $2.6641$ & $2.6276$ & $-$ & $-$ & $-$ & $-$ & $-$ & $-$  \\ \hline
3 & $3.0751$ & $3.0185$ & $-$ & $-$ & $-$ & $-$ & $-$ & $-$ \\ \hline
4 & $3.2465$ & $3.1839$ & $3.1917$ & $-$ & $-$ & $-$ & $-$ & $-$ \\ \hline
5 & $3.4029$ & $3.3358$ & $3.3382$ & $-$ & $-$ & $-$ & $-$ & $-$ \\ \hline
8 & $3.8102$ & $3.7339$ & $3.7254$ & $3.7440$ & $-$ & $-$ & $-$ & $-$ \\ \hline
10 & $4.0458$ & $3.9655$ & $3.9521$ & $3.9654$ & $-$ & $-$ & $-$ & $-$ \\ \hline
15 & $4.5559$ & $4.4688$ & $4.4469$ & $4.4513$ & $4.4698$ & $-$ & $-$ & $-$ \\ \hline
18 & $4.8241$ & $4.7341$ & $4.7087$ & $4.7094$ & $4.7240$ & $-$ & $-$ & $-$ \\ \hline
19 & $4.9087$ & $4.8178$ & $4.7915$ & $4.7911$ & $4.8046$ & $-$ & $-$ & $-$ \\ \hline
20 & $4.9913$ & $4.8996$ & $4.8723$ & $4.8709$ & $4.8834$ & $-$ & $-$ & $-$ \\ \hline
25 & $5.3773$ & $5.2821$ & $5.2509$ & $5.2453$ & $5.2535$ & $5.2703$ & $-$ & $-$ \\ \hline
35 & $6.0504$ & $5.9498$ & $5.9132$ & $5.9018$ & $5.9040$ & $5.9146$ & $5.9311$ & $-$ \\ \hline
39 & $6.2928$ & $6.1903$ & $6.1521$ & $6.1389$ & $6.1393$ & $6.1482$ & $6.1628$ & $-$ \\ \hline
40 & $6.3515$ & $6.2485$ & $6.2099$ & $6.1964$ & $6.1963$ & $6.2048$ & $6.2190$ & $-$ \\ \hline
45 & $6.6347$ & $6.5295$ & $6.4892$ & $6.4738$ & $6.4719$ & $6.4785$ & $6.4907$ & $6.5072$ \\ \hline
50 & $6.9030$ & $6.7955$ & $6.7537$ & $6.7368$ & $6.7333$ & $6.7382$ & $6.7488$ & $6.7636$ \\ \hline \hline
$x_c|_{{x_h}\uparrow}$ & $12.9161$ &$15.8189$ &$17.3288$ &$18.2661$ &$18.9073$ &$19.3742$ &$19.7297$ &$20.0096$ \\ \hline
\end{tabular}
\end{center}
\caption{The numerical results of the critical radius $x_c$ for different dimensions $n$ and different powers $k$,  where $b=0.0447$. A hyphen means that the corresponding black hole is forbidden by the hoop conjecture, so no critical horizon radius exists.}
\label{biao4}
\end{table}

\begin{figure}
\centering
\subfloat[Plot of the relations of $T_h$ with respect to $x_h$ at $b=0.0932$.]{\includegraphics[width=125mm]{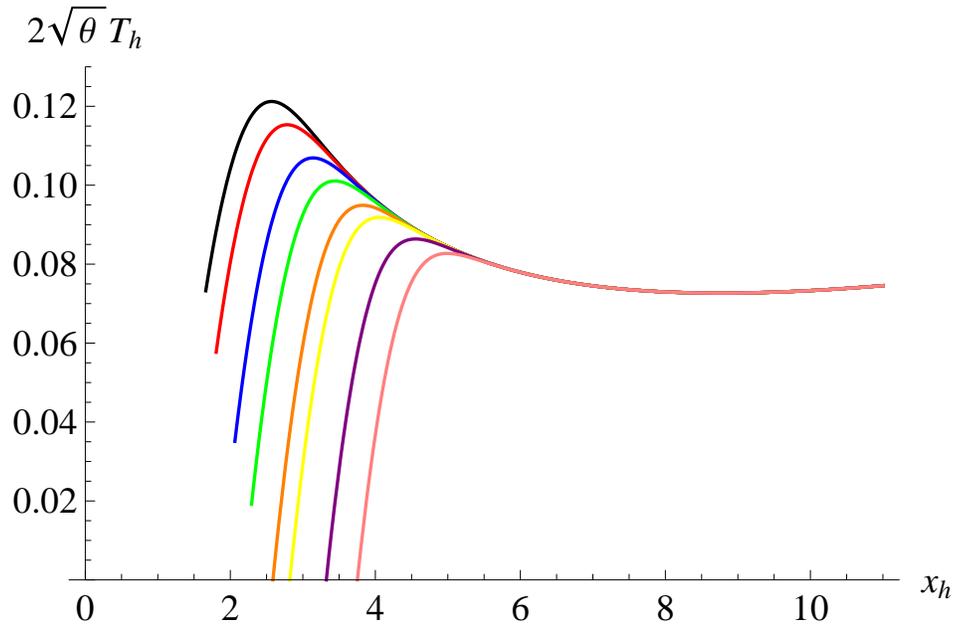}}\\
\vspace{4mm}
\subfloat[Plot of the relations of $C_p$ with respect to $x_h$ at $b=0.1222$.]{\includegraphics[width=125mm]{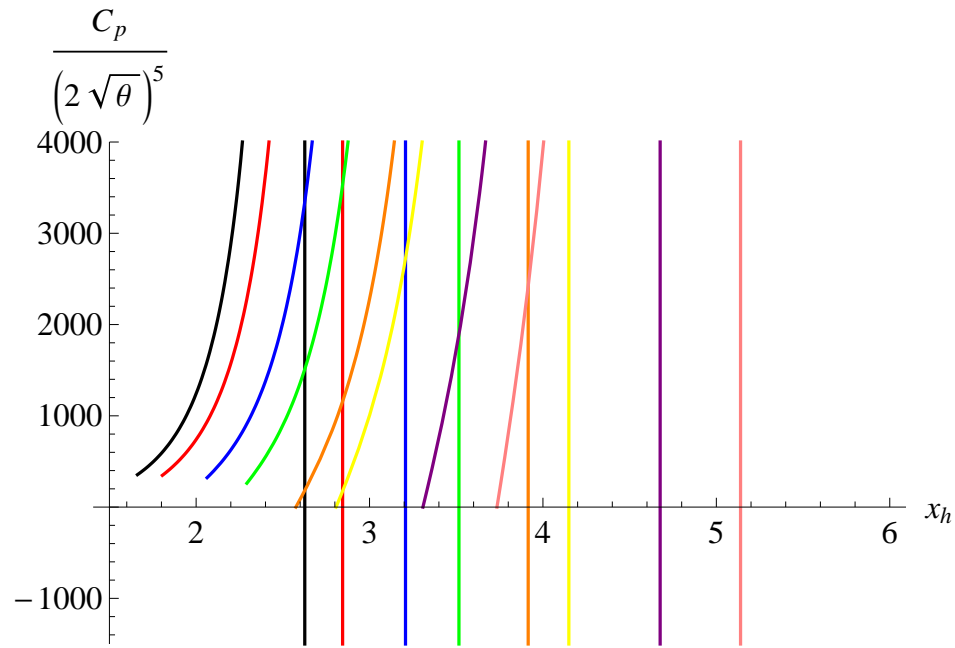}}
\caption{We take $n=7$ as an example and show how the hoop conjecture works thermodynamically for $k=0$ (\text{{\color{black}black}}), $1$ (\text{{\color{red}red}}), $3$ (\text{{\color{blue}blue}}), $5$ (\text{{\color{green}green}}), $8$ (\text{\textcolor[rgb]{1.00,0.50,0.00}{orange}}), $10$ (\text{{\color{yellow}yellow}}), $15$ (\text{\textcolor[rgb]{0.50,0.00,0.51}{purple}}), and $20$ (\text{\textcolor[rgb]{1.00,0.50,0.50}{pink}}).}
\label{hctc}
\end{figure}

It is interesting to mention how the hoop conjecture works thermodynamically. At first, we make an analysis qualitatively. If the hoop conjecture is satisfied, the extremal horizon radius of black holes is greater than the mean radius of mass distributions. As a result, the extreme black holes exist and the corresponding temperature and heat capacity equal zero. On the contrary, if the hoop conjecture is violated, the extremal horizon radius of black holes is smaller than the mean radius of mass distributions. Thus, no extreme configurations of black holes exist and the mean radius of mass distributions is just the horizon radius of the smallest black hole in order to ensure the formation of black holes, which leads of course to non-zero temperature and heat capacity for such a smallest black hole. Next, we turn to a quantitative analysis whose numerical results are plotted in Figure \ref{hctc}. In the first diagram we calculate the Hawking temperature for the cases $n=7$, $b=0.0932$, and $k=0, 1, 3, 5, 8, 10, 15, 20$, respectively. The eight cases can be classified into two groups, where the hoop conjecture is violated in the first group with $k=0, 1, 3, 5$, while it is satisfied in the second group with $k=8, 10, 15, 20$. One can see clearly that the Hawking temperature is non-zero for the smallest black holes in the first group, while it is zero for the extreme black holes in the second group. The four cases in the first group are inconsistent with the self-regularity of the noncommutative black hole~\cite{NSS}, while the other four in the second group are indeed consistent with the self-regularity. Moreover, in the second diagram of Figure \ref{hctc} we calculate the heat capacity for the cases $n=7$, $b=0.1222$, and $k=0, 1, 3, 5, 8, 10, 15, 20$, respectively. Completely following the treatment to the Hawking temperature, one can obtain the similar results, that is, the heat capacity is non-zero for the four cases in the first group in which the hoop conjecture is violated, while it is zero for the other four cases in the second group in which the hoop conjecture is satisfied. Thus, our qualitative and quantitative analyses coincide with each other. As a whole, the hoop conjecture leads, from the point of view of thermodynamics, to zero temperature and zero heat capacity for extreme black holes, which is required by the self-regularity of the noncommutative black hole.


\subsection{Entropy and the Gibbs free energy}
The entropy of this noncommutative black hole can be expressed in terms of a function of horizon radius $r_h$,
\begin{equation}
S=\int_{r_0}^{r_h}
\frac{d S}{dr_h}dr_h. \label{entr}
\end{equation}
Because the extreme black hole corresponds to zero temperature, the integration must be made from the extremal radius $r_0$ because  this radius is minimal and such a choice gives rise to a vanishing entropy for the extreme black hole. This treatment also coincides with  the third law of thermodynamics.

Now we begin with the first law of thermodynamics to calculate the entropy. The first law of thermodynamics at a constant pressure reads as $dM=T_h dS$, and alternatively $dM=\frac{\partial M}{\partial r_h} dr_h$ from the point of view of eq.~(\ref{mas}). As a result, the integrand of eq.~(\ref{entr}) takes the form,
\begin{equation}
\frac{d S}{dr_h}=\frac{1}{T_h}\frac{\partial M}{\partial r_h}.\label{sd}
\end{equation}
Again considering eqs.~(\ref{mas}) and (\ref{tem11}), we derive the left hand side of eq.~(\ref{sd}) multiplied by a suitable normalization factor,
\begin{equation}
\left(\frac{1}{2\sqrt{\theta}}\right)^{n-3}\frac{d S}{dr_h}=\frac{(n-2)\omega \Gamma\left(\frac{n+k-1}{2}\right)x_h^{n-3}}{4\gamma\left(\frac{n+k-1}{2},x_h^2\right)}. \label{Sderi}
\end{equation}
Substituting eq.~(\ref{Sderi}) into eq.~(\ref{entr}), and then integrating by parts, we obtain
\begin{equation}
\frac{S}{\left(2\sqrt{\theta}\right)^{n-2}}=\frac{\omega \Gamma\left(\frac{n+k-1}{2}\right) x_h^{n-2}}{4\gamma\left(\frac{n+k-1}{2},x_h^2\right)}+ \Delta S, \label{entr0}
\end{equation}
where $\Delta S$ reads as
\begin{equation}
\Delta S=\frac{\omega \Gamma\left(\frac{n+k-1}{2}\right)}{4}\left\{\int_{x_0}^{x_h}\frac{z^{n-3}G(n,k;z)}{\gamma\left(\frac{n+k-1}{2},z^2\right)}dz -\frac{x_0^{n-2}}{\gamma\left(\frac{n+k-1}{2},x_0^2\right)}\right\}. \label{entr1}
\end{equation}

We can verify that the entropy turns out to be the {\em standard} Bekenstein-Hawking entropy $S=\omega r_h^{n-2}/4$ under the commutative limit $\theta\rightarrow 0$ or the large horizon radius $r_h$ limit.


Now we turn to the Gibbs free energy that is defined by
\begin{equation}
G:=M(r_h)-T_h (r_h)S(r_h),\label{free}
\end{equation}
where $M(r_h)$, $T_h (r_h)$, and $S(r_h)$ are given by eq.~(\ref{mas}), eq.~(\ref{tem11}), and eqs.~(\ref{entr0}) and (\ref{entr1}), respectively.
Incidentally, under the commutative limit $\theta\rightarrow 0$ or the large horizon radius $r_h$ limit, the Gibbs free energy tends to the commutative formula~\cite{RD},
\begin{equation}
G \rightarrow \frac{\omega}{16 \pi}\left(r_h^{n-3}-\frac{r_h^{n-1}}{l^2}\right).\label{free1}
\end{equation}
Note that the above limits of entropy and Gibbs free energy show the consistency of our noncommutative generalization.

\begin{figure}
\centering
\subfloat[Plot of the relation of $G$ with respect to $x_h$,  where the extremal horizon radius is $x_0=2.0299$, the critical radii are $x_c=3.0185$ and $x_c|_{{x_h}{\uparrow}}=15.8189$. When $x_h > x_g =22.40872$, the Gibbs free energy is negative. ]{\includegraphics[width=125mm]{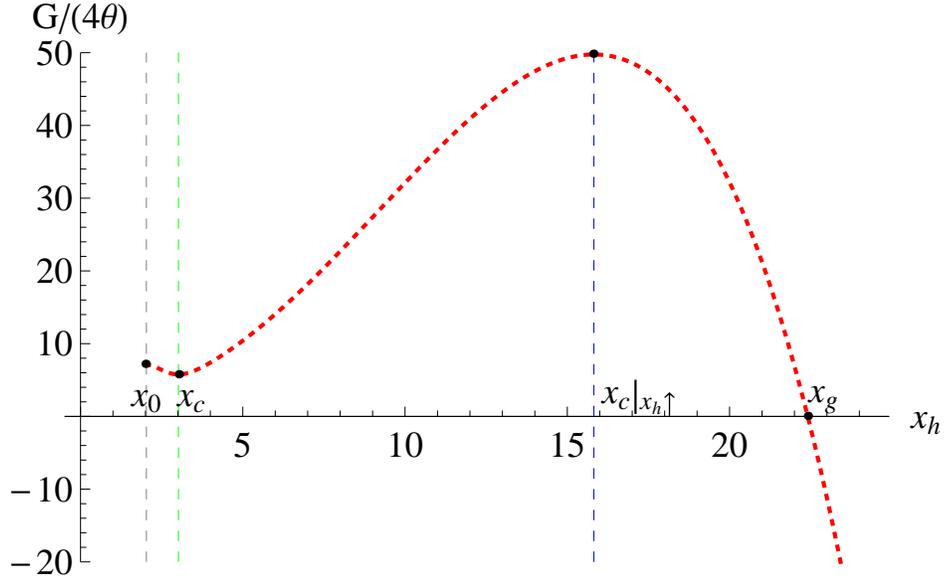}}\\
\subfloat[Plot of the relation of $G$ with respect to $T_h$. It is called the {\em swallowtail} picture, where the four dots $A$, $B$, $C$ and $D$  correspond to the horizon radii (temperature) $x_0=2.0299$ ($2\sqrt{\theta}T_h=0$), $x_c=3.0185$ ($2\sqrt{\theta}T_{max}=0.05053$), $x_c|_{{x_h}{\uparrow}}=15.8189$ ($2\sqrt{\theta}T_{min}=0.02012$), and $x_g=22.40872$ ($2\sqrt{\theta}T_{HP}=0.02135$), respectively.]{\includegraphics[width=125mm]{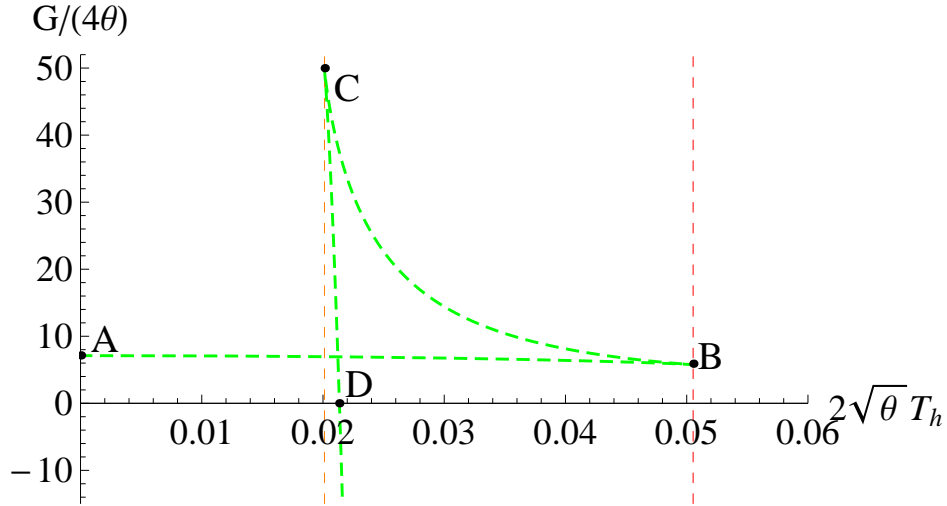}}
\caption{Plots of the relations of $G$ with respect to $x_h$ and $T_h$, respectively. We set $b=0.0447$,  $n=5$,  and $k=3$ in the both diagrams above.}
\label{tu8}
\end{figure}

As the entropy cannot be integrated analytically, which results in the fact that eq.~(\ref{free}) cannot be written in an explicit form, so we adopt the numerical method to analyze the Gibbs free energy. Eq.~(\ref{free}) is plotted in Figure \ref{tu8}, where the relevant parameters are set and can be seen in its caption. Here we point out the important value of horizon radii $x_g$ at which the Hawking-Page phase transition occurs. When $x_h=x_g$, the Gibbs free energy vanishes at the Hawking-Page temperature $T_{HP}$. When $x_h > x_g$, the Gibbs free energy becomes negative for the temperature range $T_h > T_{HP}$, indicating a stable black hole which is shown in diagram (a) of Figure \ref{tu8}. Moreover, the {\em swallowtail}  structure of the Gibbs free energy as a function of temperatures at a constant pressure is depicted in diagram (b) of Figure \ref{tu8}.

Comparing Figure \ref{tu2}, Figure \ref{tu3}, and Figure \ref{tu8}, we can separate the configurations of the noncommutative black hole into the following two classes in terms of the critical $b$-parameter.
\begin{enumerate}
\item When $b < b_c$, there exist {\em multiple} configurations about this noncommutative black hole.
\begin{itemize}
  \item If $x_h =x_0$, which means that the horizon radius shrinks to the extremal horizon radius, the black hole is in the extreme configuration. The temperature vanishes, indicating that this black hole is in the {\em frozen} state.
  \item If $x_0 < x_h < x_c$, which means that the black hole is in the near-extreme configuration, the temperature is $0 < 2\sqrt{\theta}T_{h} < 2\sqrt{\theta}T_{max}$ and the heat capacity $C_p >0$, indicating that the black hole is locally stable.
  \item If $x_h=x_c$, the temperature approaches a local maximum $2\sqrt{\theta}T_{max}$ and the Gibbs free energy drops to a local minimum. The heat capacity is divergent, which implies that the phase transition occurs.
  \item If $x_c < x_h < x_c|_{x_{h}\uparrow}$, the temperature presents a dropping trend and the Gibbs free energy presents an increasing trend. The heat capacity is negative $C_p <0$, indicating that the black hole is locally unstable.
  \item If $x_h = x_c|_{x_{h}\uparrow}$, the temperature drops to a local minimum $2\sqrt{\theta}T_{min}$ and the Gibbs free energy approaches a local maximum. The heat capacity is divergent, which implies that the phase transition occurs again.
  \item If $x_c|_{x_{h}\uparrow} < x_h < x_g$, the temperature is increasing and the Gibbs free energy is dropping. The heat capacity is positive $C_p >0$, indicating that the black hole is locally stable.
  \item If $x_h = x_g$, the Gibbs free energy is equal to zero and the temperature is equal to the Hawking-Page temperature $T_{HP}$. The first order Hawking-Page phase transition occurs between the thermal radiation and the large black hole.
  \item If $x_h > x_g$, the temperature is continuously increasing and the Gibbs free energy is negative. This black hole is in a stable state with a large radius and a high temperature.
\end{itemize}
\item When $b \geq b_c$, the extreme black hole still exists and the heat capacity is positive, implying that this black hole is locally stable. Once the horizon radius approaches $x_g$ which corresponds to the vanishing Gibbs free energy, the Hawking-Page phase transition occurs. Hence, this black hole is in a stable state with a high temperature.
\end{enumerate}

\subsection{Equation of state}
The results in the above two subsections, including the results of extreme black holes, are based on the case in which the parameter $b$ is fixed. In light of eqs.~(\ref{pres}) and (\ref{daihuan1}), we can see that the pressure parameter $4\theta P$ is proportional to the  square of the parameter $b$. So the above analysis is made in the isobaric process.

Now we adopt an alternative way, i.e. the isothermal process, to discuss the thermodynamic features of this noncommutative black hole. As a {\em priori} choice, we proceed to study the equation of state $P=P(V,T_h)$ for the noncommutative black hole in which the thermodynamic volume $V$ can be expressed as the function of the horizon radius $x_h$ from eqs.~(\ref{pres}), (\ref{daihuan1}) and (\ref{mas11}),
\begin{eqnarray}
V=\left(\frac{\partial M}{\partial P}\right)_{S}
=\frac{\left(2\sqrt{\theta}\right)^{n-1}\Gamma\left(\frac{n+k-1}{2}\right)}{\gamma\left(\frac{n+k-1}{2},x_h^2\right)}\frac{\omega}{n-1}x_h^{n-1}.
\end{eqnarray}
With the help of the eq.~(\ref{tem11}), we obtain the equation of state,
\begin{equation}
4\theta P=\frac{(n-1)(n-2)}{n-1-G(n,k;x_h)}\left[\frac{\sqrt{\theta}T_h}{2x_h}-\frac{n-3-G(n,k;x_h)}{16\pi x_h^2}\right]. \label{state}
\end{equation}
Under the commutative limit $\theta\rightarrow 0$, eq.~(\ref{state}) turns back to the known formula \cite{RD,GKM},
\begin{eqnarray}
P \rightarrow \frac{(n-2)T_h}{4r_h}-\frac{(n-2)(n-3)}{16\pi r_h^2}. \label{state1}
\end{eqnarray}

It seems that the equation of state is divergent at $x_*$ that is the solution of $n-1-G(n,k;x_*)=0$. However, according to eq.~(\ref{range}) we know $ x_* < x_0 < {\tilde x}$. This implies that such a divergence of the equation of state can be avoided since this inequality ensures that the horizon radius of black holes is always larger than $x_*$.

\begin{figure}
\centering
\subfloat[$2\sqrt{\theta}T_h=0.1000$, $0.08164$, $0.0625$, $0.04869$, $0.0150$, respectively, from top to bottom, where $n=5$  and $k=3$. The blue dashed curve that corresponds to $2\sqrt{\theta}T_{c1}=0.08164$ is the critical curve and the red dashed curve that corresponds to $2\sqrt{\theta}T_{c2}=0.04869$ is the other critical curve. When $2\sqrt{\theta}T_{c2} < 2\sqrt{\theta}T_h < 2\sqrt{\theta}T_{c1} $, the Maxwell equal area law maintains, but  it fails when $2\sqrt{\theta}T_h$ is above $2\sqrt{\theta}T_{c1}=0.08164$ and below $2\sqrt{\theta}T_{c2}=0.04869$. Note that the black curve is associated with the low temperature $2\sqrt{\theta}T_h =0.0150 < 2\sqrt{\theta}T_{c2}$, where the part of negative pressure is forbidden by the AdS background.]{\includegraphics[width=125mm]{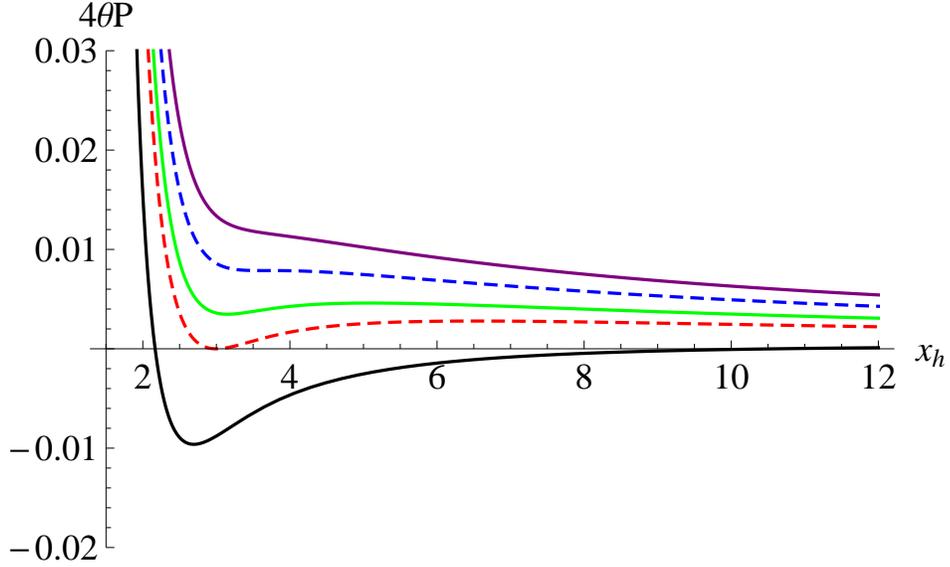}}\\
\subfloat[$2\sqrt{\theta}T_h=0.0625$, where $n=5$ and $k=3$. Here the green curve of diagram (a) is amplified in a small region.]{\includegraphics[width=125mm]{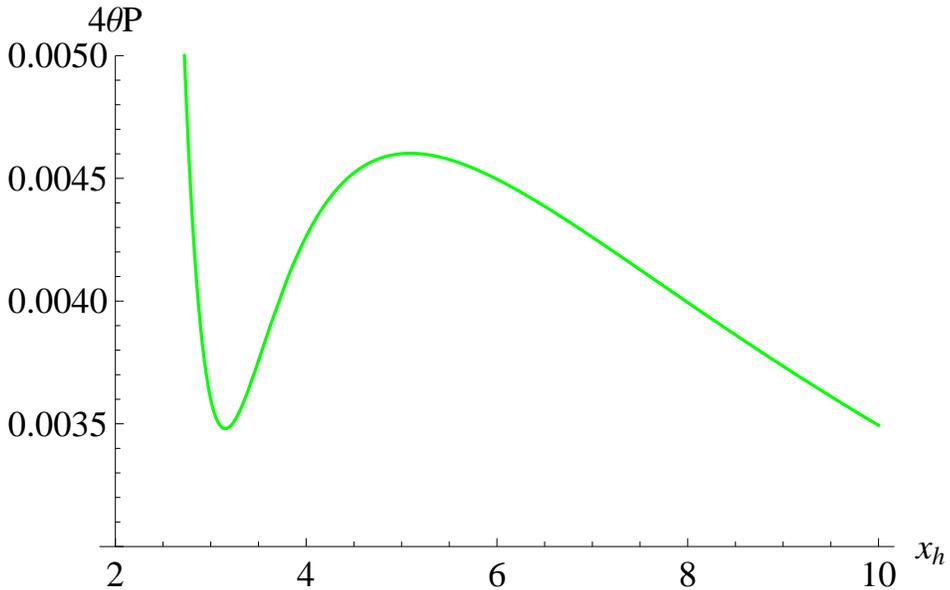}}
\caption{Plots of the relations of $P$ with respect to $x_h$.}
\label{tu4}
\end{figure}

The equation of state described by eq.~(\ref{state}) is plotted in Figure \ref{tu4}, where we can see that the Maxwell equal area law maintains for the noncommutative black hole when the Hawking temperature satisfies the inequality $T_{c2} \leq T_h < T_{c1}$, but fails when the Hawking temperature goes above the critical point $T_{c1}$ or below the other critical point $T_{c2}$. In addition, for the commutative black hole, precisely speaking, for the pure AdS black hole, whose equation of state is depicted by  eq.~(\ref{state1}), the Maxwell equal area law does not maintain as it was known.

Finally, we make a comment that the appearance of the lower bound of temperature is only associated with the validity of the Maxwell equal area law, and the extreme black holes with vanishing Hawking temperature still exist below this temperature bound. 
When $T_h <T_{c2}$, the pressure is negative in some range of horizon radius, see, for instance, the black curve of Figure \ref{tu4}. As a negative pressure that corresponds to a positive cosmological constant is contradictory to our prerequisite that our background spacetime is the AdS with a negative cosmological constant, we thus only consider the range of  horizon radius that is related to a positive pressure.
When the temperature vanishes, i.e. $T_{h} = 0$, we can see that eq.~(\ref{state}) goes back to eq.~(\ref{extreradi}) that gives the extremal horizon radius satisfying the inequality eq.~(\ref{range}). From this point of view, when $0 < T_{h} < T_{c2}$, the black hole exists with its horizon radius from the extremal one to a larger one that is associated with $T_{c2}$, and when $0 < T_{h} < T_{c1}$, the corresponding range of black hole horizon radius is from the extremal horizon radius to infinity as expected.

\section{Summary}\label{sec4}

In Section \ref{sec2}, the noncommutativity is imposed~\cite{NSS} on the high-dimensional Schwarzschild-Tangherlini anti-de Sitter black hole in terms of the non-Gaussian smeared matter distribution~\cite{Park}, the condition for the existence of extreme black holes is derived, and the radii of extreme black holes are  obtained for different dimensions $n$ and powers $k$.

In order to make this noncommutative black hole formed, we consider the {\em hoop conjecture}, that is, the extremal horizon radius  must be larger than the matter mean radius. Under this requirement, we derive the allowed values of $k$ at a fixed dimension $n$ in  the two ranges of the parameter $b=2\sqrt{\theta}/l$, see Table \ref{biao1} and Table \ref{biao2}. In particular, we indicate that the Gaussian smeared matter distribution is not applicable for the 6- and higher-dimensional Schwarzschild-Tangherlini anti-de Sitter black holes. Moreover, from the point of view of thermodynamics, the {\em hoop conjecture} ensures that the formed smallest black hole (the extreme black hole) has zero temperature and zero heat capacity, which coincides with the self-regularity of the noncommutative black holes~\cite{NSS} as expected.

In Section \ref{sec3}, the thermodynamic quantities of this noncommutative black hole are calculated, such as the Hawking temperature, heat capacity, entropy, Gibbs free energy and equation of state, in particular, the phase transition is analyzed in the isobaric process in detail. It is found that there exist two phase transitions: the first happens from a locally stable phase to a locally unstable one at a small horizon radius $x_c$, and the second phase transition occurs from a locally unstable phase to a locally stable one at a large horizon radius $x_c|_{{x_h}\uparrow}$. When the parameter $b$ is gradually increasing, these two locations $x_c$ and $x_c|_{{x_h}\uparrow}$ where the phase transitions occur are close to each other, and no phase transition will occur after $b$ is equal to and larger than the critical value $b_c$, resulting in this black hole being in a locally stable configuration. Table \ref{biao4} and Table \ref{biao3} show clearly that the critical radius $x_c$ is close to the extremal radius $x_0$, indicating that the first phase transition occurs at the near-extremal region, while the second happens at a large horizon radius. The two tables\footnote{In fact, we have calculated all numerical data when $n$ is from 4 to 11 and $k$ from $0$ to 50. Only one third of the data is put to the two tables because it is enough for us to see the tendency of the critical radius and the extremal radius.} also show that an anomalous trend of critical radii exists in the first phase transition for the 6- and higher-dimensional black holes, but no such an anomaly exists in the second phase transition.  On the other hand, our analysis of the entropy and Gibbs free energy indicates that the Gibbs free energy is negative when the horizon radius is becoming large or the noncommutative parameter is going to zero, i.e. the commutative black hole is locally stable.

Moreover, for the isothermal process, the equation of state described by eq.~(\ref{state}) reveals that the Maxwell equal area law maintains for the noncommutative black hole when the Hawking temperature satisfies the inequality $T_{c2} \leq T_h < T_{c1}$, but fails when the Hawking temperature goes above the critical point $T_{c1}$ or below  the other critical point $T_{c2}$.

At last, we point out that the reason that the noncommutative black hole has the new characteristics mentioned above originates solely from the specific mass distribution (eq.~(\ref{masfenbu})) together with its related formulation of the extremal radius $x_0$ (eqs.~(\ref{extreradi})-(\ref{tezheng})).  It can be seen clearly from eq.~(\ref{masfenbu}) that  $m(x)$ goes to the limit $M$ when $x$ is becoming large, where $x:={r}/({2\sqrt{\theta}})$. That is, the behavior of the noncommutative black hole is uniquely determined by the property of $m(x)$ together with its induced $G(n,k;x)$ at a small $x$ close to the extremal horizon radius.

\section*{Acknowledgments}
Y-GM would like to thank
W. Lerche of PH-TH Division of CERN for kind hospitality.
This work was supported in part by the National Natural
Science Foundation of China under grant No.11175090 and
by the Ministry of Education of China under grant No.20120031110027.
At last, the authors would like to thank the anonymous referee for the helpful comment that indeed greatly improves this work.

\section*{Appendix}


The lower incomplete gamma function $\gamma(a,x)$ is defined by
\begin{equation}
\gamma(a,x):=\int_0^{x}t^{a-1}e^{-t}dt, \tag{A1}
\end{equation}
where $x>0$ and $a >0$.
Its asymptotic behaviors take the forms,
\begin{equation}
\gamma(a,x) \rightarrow \frac{x^a}{a} \qquad \text{if} \qquad x\rightarrow 0, \tag{A2}
\end{equation}
\begin{equation}
\gamma(a,x) \rightarrow \Gamma(a) \qquad \text{if} \qquad x\rightarrow \infty, \tag{A3}
\end{equation}
where $\Gamma(a):=\int_0^{\infty} t^{a-1} e^{-t} dt$.

We have introduced a new function $G(n,k;x)$ which is defined by eq.~(\ref{tezheng}) for $x>0$, $n \geq 4$, and $k$ is a non-negative integer. It can be checked that the first order derivative of $G(n,k;x)$ with respect to $x$ is negative, i.e. $G^{\prime}(n,k;x)<0$ for $x>0$,  which implies that $G(n,k;x)$ is monotone decreasing for fixed $n$ and $k$ at the region $x>0$.

Based on the asymptotic behaviors of $\gamma(a,x)$,  now we analyze $G(n,k;x)$. When $x$ is small, we make an expansion in Taylor series for $G(n,k;x)$,
\begin{equation}
G(n,k;x)=(n+k-1)+\left(\frac{4}{n+k+1}-2\right)x^2+\frac{8(n+k-1)}{(n+k+1)^2 (n+k+3)}x^4+\mathcal{O}(x^6), \tag{A4}
\end{equation}
from which its asymptotic behavior is
\begin{equation}
G(n,k;x)\simeq (n+k-1)+\left(\frac{4}{n+k+1}-2\right)x^2. \tag{A5}
\end{equation}
Moreover, when $x\rightarrow \infty$, the asymptotic behavior of $G(n,k;x)$ reads
\begin{equation}
G(n,k;x) \rightarrow 0.  \tag{A6}
\end{equation}


\end{document}